\documentclass[11pt]{article}

\usepackage{amssymb,amsmath}

\usepackage{amsmath}
\usepackage{amssymb}
\usepackage{amsthm}
\usepackage{url}

\newcommand\x{{\bf x}}

\newcommand\bfa{{\bf a}}

\newcommand\bff{{\bf f}}

\newcommand\zero{{\bf 0}}
\newcommand\cc{{\mathbb C}}

\theoremstyle{definition}

\begin{document}

\title{Evaluating polynomials in several variables and their derivatives
 on a GPU computing processor\thanks{This
material is based upon work supported by the National Science Foundation
under Grant No.\ 1115777.}}
\author{Jan Verschelde and Genady Yoffe \\
Department of Mathematics, Statistics, and Computer Science \\
University of Illinois at Chicago \\
851 South Morgan (M/C 249) \\
Chicago, IL 60607-7045, USA \\
\texttt{jan@math.uic.edu, gyoffe2@uic.edu} \\
~~~~~~~\texttt{\url{www.math.uic.edu/~jan}},
\texttt{\url{www.math.uic.edu/~gyoffe}}
}

% \date{\today}
\date{31 December 2011}

\maketitle

\begin{abstract}
In order to obtain more accurate solutions of polynomial systems 
with numerical continuation methods we use multiprecision arithmetic. 
Our goal is to offset the overhead of double double arithmetic 
accelerating the path trackers and in particular Newton's method
with a general purpose graphics processing unit.
In this paper we describe algorithms for the massively parallel evaluation
and differentiation of sparse polynomials in several variables.
We report on our implementation of the algorithmic differentiation
of products of variables on the NVIDIA Tesla C2050 Computing Processor
using the NVIDIA CUDA compiler tools.

\medskip

{\em Key words and phrases.} algorithmic differentiation, 
compute unified device architecture (CUDA),
graphics processing unit (GPU),
massively parallel polynomial evaluation, 
Speelpenning product.
\end{abstract}

\section{Introduction}

The problem we consider in this paper is the efficient evaluation
of a polynomial system and its Jacobian matrix on a graphics processing 
unit (GPU), the NVIDIA Tesla C2050 Computing Processor.
For an introduction to GPU computing,
we refer to~\cite{KH10} and~\cite{NVIDIA}.
The success of general purpose GPU computing in many areas of scientific
computing is explained in~\cite{Hwu11}.

The evaluation of a polynomial system and its Jacobian matrix is a
computationally intensive stage in Newton's method to approximate an
isolated solution.  Numerical continuation methods apply Newton's method
as corrector in predictor-corrector algorithms to track paths of solutions
defined by a homotopy.  Homotopy continuation methods have led to efficient
numerical solvers of polynomial systems
(see e.g.~\cite{AG03}, \cite{Li03}, \cite{Mor87}, \cite{Wat02})
and constitute the computational engine in the emerging area of numerical 
algebraic geometry (\cite{Ley11}, \cite{SVW05}, \cite{SW05}).

Granularity issues and parallel complexity of continuation methods
for nonlinear systems are discussed in~\cite{ACW89} and~\cite{CARW93}.
If one is interested in computing {\em all} isolated solutions of a
polynomial system, then distributing path tracking jobs in a manager/worker
paradigm using message passing~\cite{MPI98} leads to very good speedups.
Such parallel implementations are in {\tt Bertini}~\cite{BHS11},
{\tt HOM4PS-2.0para}~\cite{LT09}, {\tt PHoMpara}~\cite{GKFK06},
{\tt POLSYS\_GLP}~\cite{SMSW06}, and {\tt PHCpack}~\cite{Ver99},
documented in~\cite{GV08}, \cite{LV05,LV09}, \cite{LVZh06}, \cite{VW04}, 
and~\cite{VZ06}.

For large polynomial systems in many variables and of high degrees we
have observed that (1) the cost of polynomial evaluation often dominates
the cost of linear algebra operations; and (2) the double precision in
standard hardware is often insufficient to guarantee accurate results.
When running many path tracking jobs, a couple or perhaps just one
solution path may require extended multiprecision arithmetic.
Then we want to apply parallel algorithms
to offset the extra cost of software driven arithmetic.
In analogy to speedup, we use the notion of {\em quality up}
(inspired by~\cite{Akl04}) and ask the question: given $p$ processors
(or cores) how much extra precision can we afford in roughly the
same time as a sequential run?

Because of simple memory management on shared memory multicore processing
we selected the quad double library {\tt QD~2.3.9}~\cite{HLB00}.
The ideas to achieve extended precision using hardware doubles
originate in~\cite{Dek71}, see also~\cite{Pri92}, \cite{Rum10} and~\cite{She97}.
In~\cite{VY10}, we determined experimentally that 
the cost factor in the overhead of using double double arithmetic is
around 8, coinciding with the number of cores on the workstation
we were using at that time.
Then the cost of tracking one solution path in double double arithmetic
can be compensated in a parallel multicore implementation,
thus achieving quality up.
Using techniques of algorithmic differentiation~\cite{GW08},
we extended this work in~\cite{VY11}.

This current paper describes our efforts to offset the cost of
extended precision by a parallel implementation of evaluation
and differentiation algorithms on a GPU.
Recently, quad double precision arithmetic has been made
available on GPU, see~\cite{LHL10}.
Our long term goal is use GPUs to accelerate
the solver of PHCpack~\cite{Ver99}.

Related work in algebraic computations on a GPU
are polynomial multiplication~\cite{Eme09}, \cite{MP10},
resultant~\cite{Eme10}, GCD computations~\cite{Eme11},
and solving bivariate polynomial systems~\cite{MP11}.
Parallel automatic differentiation techniques are
described in~\cite{BGKW08} and~\cite{GPGK08}.

\section{Problem Statement}

The problem we consider is the evaluation of system of polynomial
equations in several variables and all its derivatives 
(as needed in the Jacobian matrix of the system).
Let $n$ denote the number of variables.
A polynomial $f$ in $n$ variables $\x = (x_1, x_2, \ldots, x_n)$
is stored as a tuple $(C,A)$ of complex coefficients $C$ and
corresponding exponents $A$.  In multi-index notation we write $f$ as
\begin{equation}
   f(\x) = \sum_{\bfa \in A} c_\bfa \x^\bfa,
   \quad c \in \cc \setminus \{ 0 \}, 
   \quad \x^\bfa = x_1^{a_1} x_2^{a_2} \cdots x_n^{a_n}.
\end{equation}
Then a system $\bff(\x) = \zero$ is defined by a tuple of coefficients
and supports.

In our problem setup, we consider as inputs {\em sparse} polynomials,
that is: only relatively few monomials appear with nonzero coefficients,
few relative to the degree of the polynomials.  
For dense polynomials, a nested Horner scheme is recommended,
see~\cite{Koj08}.
Because of the exponential growth of the number of monomials,
dense polynomials in several variables of high degree
do not occur often in applications. 
For establishing benchmarks we consider in this paper systems
with a fixed number $k$ of variables in monomials, a fixed maximal degree $d$
up to which any of variables can appear in monomials of the system,
and  a fixed number $m$ of monomials in all polynomials.

There are three stages in the evaluation of a polynomial system
and its Jacobian matrix:
\begin{enumerate}
\item The computation of the monomial products 
      $x_{i_1}^{a_{i_1}-1} x_{i_2}^{a_{i_2}-1} \cdots x_{i_k}^{a_{i_k}-1}$
      where the exponents  $a_{i_j} \geq 1$, $j=1,2,\ldots,k$,
      and $1 \leq i_1 < i_2 < \ldots <  i_k \leq n$.
      A preprocessing step is the computation of all powers of~$x_i^j$,
      for all $i \in \{1,2,\ldots,n\}$ and $j \in \{2,\ldots,d-1\}$.

\vspace{1mm}
          
\item The evaluation of products of variables
      $x_{i_1} x_{i_2} \cdots x_{i_k}$,
      and all their derivatives.
      The product of variables $x_{i_1} x_{i_2} \cdots x_{i_k}$
      is known in~\cite{GW08} as the example of Speelpenning.
      In this stage we also multiply Speelpenning products and their derivatives with the values of monomial products
      computed in the previous stage to obtain the values of monomials and their derivatives.
 
 \vspace{1mm}
      
\item The multiplication of the coefficients with the corresponding 
          evaluated monomials,  followed by  summations of obtained products within
          hosting polynomials of the system and the Jacobian.
 
 \end{enumerate}

While the evaluation of high dimensional and high degree polynomials
is computationally intensive, the challenge for data parallelism
lies in the irregularity of the data.  In order to achieve good speedups,
we will derive regularity assumptions on the input data.

\section{Massively Parallel Algorithms}

Following the three stages (outlined in the problem statement section)
in the evaluation of a polynomial system and its Jacobian matrix,
we devote one subsection to each stage.
There are three kernels.  The first kernel corresponds to the first stage,
whereas the second kernel is described in subsections 2 and~3 below.
We explain the third kernel in the second half of the third subsection.

\subsection{Common Factor Calculation}

To evaluate a monomial $x_1^3 x_2^7 x_3^2$ and its derivatives,
we first evaluate the factor $x_1^2 x_2^6 x_3$ and then multiply
this factor with all derivatives of $x_1 x_2 x_3$.
Because $x_1^2 x_2^6 x_3$ is common to the evaluated monomial
and all its derivatives, we call $x_1^2 x_2^6 x_3$ a common factor.
This section is concerned with the evaluation of all common factors.

The kernel to compute common factors operates in two stages:
\begin{enumerate}
\item each of the first $n$ threads of a thread block computes sequentially
      powers from the 2nd to the $(d-1)$th of one of the $n$ variables;
\item each of the threads of a block computes a common factor for one 
      of the monomials of the system, just as a product of $k$ quantities 
      computed at the first stage of the kernel.
\end{enumerate}
Storing the values of the successive variables in the successive
locations of the global memory enables a coalesced reading by the 
threads of a warp of the values of the variables of the system 
from the global memory into the shared memory, 
as an input for the first stage of the kernel.

Both stages of the kernel are largely SIMT (Single Instruction Multiple
Thread) routines since at the first stage each of the non idle threads 
just performs $d-1$ multiplications, and at the second stage each
thread in each warp just performs $k-1$ multiplications.

The precomputed at the first stage powers of variables, are stored at
the shared memory of the blocks, since these powers essentially constitute
shared input data for the threads of the block while the threads are
working on the second stage of the kernel.  The powers are stored in
shared memory in a two dimensional array {\sc Powers} of complex numbers,
an $(i,j)$th element of which represents the $i$th power of the $j$th variable.
Such indexing is supposed to minimize a number of shared memory bank
conflicts at least during the first stage of the kernel, as different threads in
a warp, after computing the current power of associated to them
variables, will be writing the power values into different banks of the shared
memory.

As the threads of a block perform the second stage of the kernel, each
of them computes a product of $k$ quantities, computed at the first stage.
As a thread proceeds to each next element in a product, to know what element to access in the
shared memory array {\sc Powers}, it  would need to know which variable
and what exponent appears next in the assigned to it monomial.
The information about positions of variables and their exponents does not
change along the path tracking, and thus might be stored in constant
memory of the card. 
We reserve two arrays of unsigned chars {\sc Positions} and
{\sc Exponents} in the constant memory to represent this information. Each
element in {\sc Positions} represents a position of a variable from 0 to 
$255$ in one of monomials of the system, and the element with the same 
index in {\sc Exponents} represents the degree of this variable decreased
by one in the same monomial, giving us opportunity to work with variables 
appearing in degrees up to~255. 

We need at least about 1,000 monomials to occupy well
all the 14 multiprocessors of our card for the algorithms we consider
here, so several warps would  work on each multiprocessor simultaneously
to hide long latency operations. This and the capacity of the constant 
memory, 65,536 bytes, prescribes working
dimensions for our polynomial evaluation: those are ranging from 30 to 40.
If we want to keep $m$, the number of monomials in the polynomial,
to be equal roughly to the dimension of the system, and $k$, the number 
of variables in the monomial, about half of the dimension.
Indeed: for dimension 30 we would have 900 monomials, with a need of
$900 \times 2 \times 15 \leq 30,000$ bytes; 
for dimension 40 we would have 1,600 monomials,
with a need of $1,600 \times 2 \times 20=64,000$ bytes.

We are planning to introduce more compact encodings for storing the
positions and exponents of the variables in the constant memory so to be
working with higher dimensions. The more compact encodings
might introduce some branching for the threads of a warp, after the
decoded indexing information would be read from the constant memory into
the registers of the block, while each thread in a warp would be encoding
the actual position and exponent of the next variable’s
power, which it needs to use for its computations. 
However the computations,
which would follow encodings (the multiplications), where the
threads of a warp will join again one path of execution, are supposed to
dominate encodings in time, especially if higher precision multiplications
would be used. Thus with new ways of decoding, incorporated to store more efficiently
monomial information in the constant memory, and employed multiprecision,
we hope increase working dimensions for our implementation.

After each thread of a block computes its common factor, the successive
threads of the block conveniently write their output values (one value
per thread) into successive locations of the global memory,
thus providing a coalesced output for the kernel.

As an alternative to computing common factors in the two above stages,
one can skip precomputing powers, and assign to each thread all work,
which is necessary for computing of assigned to it common factor, to do
by itself from scratch. This could be done entirely in registers assigned
to a block, without any use of shared memory.  However this would
introduce branching in execution of threads of a warp when monomials
would have different tuples of exponents, and if one would choose that
each thread would compute all powers up to $d-1$ for participating in
its monomial variables, it would most likely cause multiple exponentiation
of the same variables by threads within warps since the same variables
tend to appear in multiple monomials of a system.  In our algorithm
powers of variables are also computed multiple times -- each block of
threads computes its own copy of the set of powers from 2nd to $d-1$ 
for all $n$ variables of the system.  

This might look as a drawback of the algorithm. 
This is not really so. For our working dimensions 
ranging from 30 to 40, and
the number of maximal cores for one multiprocessor 32, we would need
to assign at most two blocks to work on precomputing degrees if we want
to do it only once, in this case 12 of 14 multiprocessors would be idle
during precomputing powers of variables.  Also to start using the other 12
multiprocessors for the second stage of computing common factors,
we would need to write the precomputed powers into the global
memory, then to invoke a separate kernel with enough blocks to occupy all
multiprocessors, and then threads of each block of the new kernel will
access the global memory again for reading the powers of variables stored
there. Our algorithm, as an alternative to prompted by the just described
two kernels scheme additional time cost for global memory reading and
writing, introduces the additional time cost, which is illustrated well by
the following example.

Consider a system of dimension 32 with 28 monomials in each polynomial.
If we will work with blocks of 32 threads, 28 blocks of threads will
be launched. Then, in the worst case, if only one block will be occupying
one multiprocessor at a time, the execution time for our two-stages kernel
will be the same as if one block of 32 threads would be launched two times
in a row.  Thus precomputing powers, despite in fact it would be done 28
times, time-wise would take the same amount of time as it would
be done twice. Then, as within one thread block powers of all variables are
computed in parallel, for our example then precomputing degrees would
take in the worst case the same time as is needed for one core to
compute $2(d - 2)$ multiplications (variables for the common factors need
to be raised  up to the power $d-1$, which requires $d-2$ multiplications).
The degree~$d$ is in most cases not that high
(while still allowing high total degrees for monomials), 
thus multiple precomputing degrees in our two stages one
kernel algorithm in most cases would compensate for the additional
necessary global memory accesses as the powers are precomputed only once,
and most likely, even reduce the computational time for precomputing powers.

\subsection{Monomial Evaluation and Differentiation of Products of Variables}

In this section we describe the implementation of the algorithm to
evaluate a product of variables $x_{i_1} x_{i_2} \cdots x_{i_k}$ 
and all its derivatives.
We call this product of variables a Speelpenning product.

In our second kernel each thread first computes one monomial and its partial
derivatives. Secondly it multiplies the computed value of the monomial by
its coefficient in the hosting that monomial polynomial of the system,
as well as it multiplies the values of the computed derivatives of the
monomial by their coefficients in the hosting those monomial derivatives
polynomials of the Jacobian. Thus this kernel completes computing additive
terms of the polynomials of the system and the Jacobian, and the third 
last kernel only adds appearing in each polynomial terms to finish polynomial
evaluation.

A thread of the second kernel performs only $5k-4$ multiplications
and uses $k + 1$ complex double locations of shared memory
$L_1, L_2, \ldots , L_{k+1}$ and one variable in registers
to perform all the announced above work.
As was discussed in the previous section, through an example, we
obtain the partial derivatives of a monomial $\x^\bfa$
by multiplying the common factor $x_{i_1}^{a_{i_1}-1}
x_{i_2}^{a_{i_2}-1} \cdots x_{i_k}^{a_{i_k}-1}$
by the partial derivatives of $x_{i_1} x_{i_2} \cdots x_{i_k}$.
It takes $3k-6$ multiplications out of 
$5k-4$ multiplications performed by a thread
to compute partial derivatives of Speelpenning product. 
Another $k$ multiplications
are performed to multiply the common factor by the values of derivatives
of Speelpenning product to obtain monomial derivatives. One additional
multiplication is done to obtain the value of the monomial itself as a
product of a derivative
of the monomial with respect to any of participating in it variables and
the value of that variable. 
Finally a thread performs another $k+1$ multiplications
to multiply the values of the monomial and its derivatives by the
coefficients.

To obtain derivatives of Speelpenning product
a thread first stores  $x_{i_1}$ in the location $L_2$.
Then it  computes sequentially, by $k-2$ multiplications, 
the $k-2$ forward products  
$x_{i_1} x_{i_2}$,  $x_{i_1} x_{i_2} x_{i_3}$, $\ldots$ ,
$x_{i_1} x_{i_2} x_{i_3} \cdots  x_{i_{k-1}}$, 
for each new $r$ ranging from 1 to $k-2$ obtaining the product
$x_{i_1} x_{i_2} x_{i_3} \cdots x_{i_{r+1}}$  
as $(x_{i_1} x_{i_2} x_{i_3} \cdots  x_{i_r}) x_{i_{r+1}}$ 
and storing the newly obtained forward product into location~$L_{r+2}$.
Eventually the locations $L_3,\ldots,L_k$ are filled with the
$k-2$ obtained forward products. 
Note that at this point the location $L_k$ contains the derivative 
of the Speelpenning product with respect to~$x_{i_k}$. 
In registers of the block we keep the only 
complex double variable $Q$ to store the current backward product.
We initialize~$Q$ with~$x_{i_k}$.
A thread computes the derivative of the Speelpenning product  
with respect to $x_{i_{k-1}}$ at $L_{k-1}$ 
by multiplying stored in that location the forward product
$x_{i_1} x_{i_2} x_{i_3} \cdots x_{i_{k-2}}$ by the current value
of $Q$, which is~$x_{i_k}$. 

In the next $k - 3$ steps,
each of which consists of two multiplications, 
we compute partial derivatives of the Speelpenning product 
with respect to $x_{i_2}, x_{i_3}, \ldots ,x_{i_{k-2}}$,
and store the computed values in locations $L_2, L_3, \ldots , L_{k-2}$.
At the $r$th step, as $r$ ranges from 1 to $k-3$, 
the $Q$ represents the backward product 
$x_{i_k} x_{i_{k-1}} \cdots x_{i_{k-r}}$.
At the $r$th step we first update the value of
$Q$, accordingly to its above definition,
by one multiplication as
$Q = Q \times x_{i_{k-r}}$.
The second multiplication updates the shared memory location
$L_{k-r-1}$ as $L_{k-r-1} = L_{k-r-1} \times Q$,
so to obtain in this location the partial derivative of  Speelpenning
product with respect to $x_{i_{k-r-1}}$ as a product of previously stored
there forward product $x_{i_1} x_{i_2} x_{i_3} \cdots x_{i_{k-r-2}}$
times the current backward product 
$x_{i_k} x_{i_{k-1}} \cdots x_{i_{k-r}}$. 

Finally we obtain the last yet
not obtained partial derivative of
Speelpenning product with respect to $x_{i_1}$ at
$Q$, by the product $Q = Q \times x_{i_2}$ 
and store the obtained value at the shared memory location $L_1$.

The above procedure prescribes to a thread to perform 
$k-2$ multiplications to obtain the forward products, 
$k-2$ multiplications to obtain the backward products,
and $k-2$ multiplications of backward and forward products. 
Thus indeed, the total number of multiplications for obtaining 
all derivatives of the Speelpenning product equals~$3k-6$.

Now a thread computes monomial derivatives
in locations $L_1, L_2, \ldots , L_k$ 
by multiplying stored in these locations values of derivatives 
of Speelpenning product by the common factor computed in
the first kernel. 
Then it computes the value of the monomial itself as the
product of its derivative with respect to $x_{i_k}$, 
stored in $L_k$ times the value of $x_{i_k}$.
It stores the computed monomial value at $L_{k+1}$.
Finally it multiplies each of the values stored in
$L_1, L_2, \ldots , L_{k+1}$,
i.e.: the values of the monomial and its
derivatives, by the corresponding coefficients.

As $k$ -- the number of variables in a monomial -- is the same for all
monomials of the system,each thread of the second kernel will go through
the same path of execution for the entire list of instructions of the
kernel, which largely amounts to $5k-4$ complex double multiplications. 
Thus all 32 threads within each warp will be indeed doing all the
prescribed work for the assigned to them 32 monomials in a parallel
fashion on an available multiprocessor during the execution.

We close this subsection with some memory considerations.
Denote by $B$ the block size, the number of threads in a block.

In addition to the fast access space of $B(k+1)$ locations equally
divided between threads of a block for storing their intermediate results,
as the threads proceed along the kernel, we reserve in shared memory of a
block a space for values of all variables of the system. The values of the
variables are subject shared use of the threads of each block, as the same
variables appear in different monomials.
Thus, provided values of successive variables are stored in successive
locations of global memory, and working with $n=32$, $k=16$, $B=32$,
we would need to access global memory only once by all threads of a block
simultaneously, as each thread would request a value of one variable, to
download the values of all 32 variables into the shared memory of the
block for their further common use by all threads of the block. 

At the same time, if shared memory would not be used for storing values of
variables, each thread would need to access global memory at least 16
times to get the values of all appearing in its monomial variables.
The shared memory capacity allows us to apply the above algorithm of the
second kernel for our working dimensions 
ranging between 30 and 40 as well as for some larger dimensions. 
We also could increase precision from double to double double
and still work with dimensions up to 70, as long as $k$ is less or equal
than a half of dimension.
Indeed, each thread would need for treating its monomial 
$k+1$ complex double double locations, thus 
\begin{eqnarray*}
   & & (n/2+1) \times 2 \times \mbox{\tt sizeof(double double)} \\
   & & \leq (70/2+1) \times 2 \times 16 = 1152
\end{eqnarray*}
bytes in shared memory.  To treat 32 monomials by a block of 32 threads 
we would need then at most $32 \times 1152=36864$ bytes of shared memory.
Adding to this 
\begin{eqnarray*}
  & & n \times \mbox{\tt sizeof(complex double double)} \\
  & &  \leq 70 \times 2 \times \mbox{\tt sizeof(double double)} \\
  & & = 70 \times 2 \times 16 = 2240
\end{eqnarray*}
bytes in shared memory for storing values of the variables, 
we are still $(49,152 - (36,864+2,240) ) > 10,000$ 
bytes below the maximal capacity of the shared memory of a block.

Another important note about the memory management is  that the array
{\sc Positions} in constant memory, which contains positions indexes of
variables in the monomials, and used in the first kernel, is used in this
kernel as well, as threads are determining what variable in the shared
memory to access as they need to perform each new multiplication while
updating their forward and backward products.

\subsection{Multiplication of Evaluated Monomials with Coefficients and Summation of Terms}

In the third stage the evaluated monomials first are multiplied
with their coefficients in the polynomials of the system or the Jacobian.

The coefficients are stored in the
global memory, since the capacity of the constant memory is exhausted by
the variables positions indexes and variables exponents information.
As we multiply monomials and their derivatives by the coefficients, we need
to read the values of coefficients from the global memory fast.
The total number of monomials in the system is $n \times m$.
For mapping purposes all the
monomials are ordered in a sequence $S_m$ of length $n \times m$.
For instance the monomials in $S_m$ might be ordered as following:
first $m$ elements of the sequence are the monomials of the first
polynomial, the next $m$ elements are the monomials of the second
polynomial, and so on.
The coefficients  are
stored during entire path tracking  in an
array {\sc Coeffs} of length $n \times m \times (k+1)$, 
which is the total number of monomials
in the system and its Jacobian. 
The coefficients in {\sc Coeffs} are stored in the following order:
\begin{itemize}
\item The first element of {\sc Coeffs}
      is the coefficient of the derivative of the
      first monomial in $S_m$ with respect to its first variable;
\item the second element of Coeffs is the coefficient of the derivative 
      of the second monomial in $S_m$ with respect to its first variable,
      and so on until 
\item the $nm$th element of {\sc Coeffs}, which is the coefficient
      of the derivative of the last monomial in $S_m$ with respect
      to its first variable. 
\item The next $n\times m$ elements of {\sc Coeffs} are the
      coefficients of the derivatives of monomials from $S_m$,
      with respect to the monomials second variables,
      also listed in accordance with order in $S_m$.  
\end{itemize}
The portions of $nm$ coefficients come in a similar manner until
the $k$th portion of $nm$ coefficients, in which are stored, in order
inherited from $S_m$, the coefficients of monomial derivatives with
respect to the monomials $k$th (last) variables.
The last $(k+1)$th portion of the $nm$ coefficients contains actually the
coefficients of the system in order prescribed by order in~$S_m$.
With this way of storing coefficients, if $i$th thread of the second kernel
is in charge of $i$th monomial in $S_m$ for each $i = 1,2,\dots,nm$,
we largely obtain a coalesced access within warps, as threads of a warp
prescribed simultaneously to access the coefficients of their monomials
or the $j$th, $j \in \{ 1,2,\ldots,k\}$, 
derivative's coefficients of their monomials in {\sc Coeffs}.

After multiplying the monomial and their derivatives values by coefficients,
which is the last computational step of the second kernel, it is just left
to add the corresponding computed additive terms to obtain the values of
the polynomials
of the system and of the Jacobian. If the size of a thread block used for
the execution of the second kernel is smaller than~$m$, then monomials of
each polynomial of the system are treated by multiple blocks of the second
kernel.
In this case, even if some of the involved summations are done yet by the
threads of the second kernel, it is necessary to launch another kernel to
combine partial sums which are obtained by different blocks of the second
kernel, which are working on monomials of the same polynomials. The
situation, when the size of a thread block of the second kernel is less
than m, is very common for our working dimensions: we try to keep the
block size of the second kernel equal to 32,
because of described above shared memory limited capacity considerations,
on the
other hand, we are willing to work with higher dimensions, ranging from 50
to 70, while we want to keep $m \approx n$. Also, computing partial sums
for polynomials of the Jacobian by threads of the second kernel would
involve branching in execution paths of the threads within warps, as
different subsets of variables appear in monomials treated by different
threads within a warp.
Because of the above reasons we decided to introduce a third kernel, which
would perform all involved summations, so to complete obtaining the values
of the polynomials, as all multiplicative operations are done by the first
two kernels.

Each thread of the third kernel sums additive terms of one of $n^2+n$
polynomials of the combined set of polynomials of the system and the
Jacobian matrix.
To make each thread to go through the same execution path, all what we assign
to each thread to do during the execution of the kernel is to add exactly
$m$ terms.
Thus, if a thread computes the value of the derivative of the $p$th
polynomial with respect to~$x_i$, and a $j$th monomial in the $p$th 
polynomial does not contain $x_{i}$, the thread which computes the derivative
of the $p$th polynomial with respect to $x_{i}$, at the $j$th step
does add to its current partial sum zero -- the zero monomial derivative,
which we probably never would add in a CPU execution. 
To ensure this, without introducing any if statements, 
the output array of the second kernel in the global memory
along with its meaningful $nm(k+1)$ locations (the number of monomials
and monomial derivatives of the system) contains also 
$(n^2+n) m - n m (k+1)$ locations, the values at which are originally 
set and kept to store zero values along the entire path tracking. 
These zero locations represent the zero monomial
derivatives as in the described above situation. 
We also wish that the threads within warps of the third kernel 
for each step~$j$, $j = 1, 2 \ldots m$ 
would perform a coalesced reading of the input data entries.
To allow coalesced reading of the values of monomials and their derivatives
by the threads of the third kernel, and to introduce the
$(n^2+n) m - n m (k+1)$ zero monomial derivatives,
the output of the second kernel is stored in the global memory
in array {\sc Mons} in the format we explain next.

The size of the array {\sc Mons} is $(n^2+n) m$, 
representing the terms in $n^2+n$ summations, $m$ terms each. 
The first $n^2+n$ elements of the array represent the first terms 
in each of $n^2+n$ summations (polynomials).
In particular, these first $n^2+n$ elements are: 
the first $n$ elements are the first monomials of the polynomials of 
the system, the second $n$ elements are the derivatives of the first 
monomials with respect to $x_1$, the third $n$ elements are the 
derivatives of the first monomials with respect to $x_2$, and so on until
the $(n+1)$th $n$ elements, which are the derivatives of the first monomials
with respect to~$x_n$. The second $n^2+n$ elements represent the second
terms in each of $n^2+n$ summations, and again the first $n$ elements of
them represent the second monomials of the polynomials of the system,
and the next $n^2$ elements represent the partial derivatives of the 
second monomials of the system, listed in the same order as are listed 
the derivatives of the first monomials. 
In general the $j$th $n^2+n$ elements represent $j$th monomials 
of the polynomials of the system and their partial derivatives listed 
in the same order as listed the first monomials of the system and their 
partial derivatives at the first $n^2+n$ elements of the array.

For simplicity in this description we assumed that the number $B$
of threads in a block, the block size, divides $n^2 + n$.
Now if we launch $(n^2+n)/B$ blocks, 
with a thread $t = BlockId \times B+ThreadId$ computing the sum: 
$\sum_{j=0}^{m-1} \mbox{\sc Mons}[t + j(n^2+n)]$,
the obtained sums will represent the values
of polynomials of the system and of the Jacobian, while access
to the elements of {\sc Mons} will be coalesced within warps 
at each step $j = 0,1,\ldots,m-1$ of the summation. 
To create the array {\sc Mons} in such a format, 
we had to make the threads of the second kernel to output 
the values of monomials and their derivatives not in a coalesced way.
However there was a tradeoff:
\begin{itemize}
\item either to make the output of the second kernel coalesced and then 
      the input of the third kernel could not be accessed in a coalesced way, 
\item or as we chosen to provide ability for the threads of the third kernel
      to read the input data in a coalesced way, in a price of not 
      coalesced writing of the output of the second kernel.
\end{itemize}

\section{Computational Experiments}

Our computations are done on a HP Z800 workstation,
running Red Hat Enterprise Linux Workstation release~6.1.
The CPU is an Intel Xeon X5690 at 3.47~Ghz.  
The processor clock of the NVIDIA Tesla C2050 Computing Processor
runs at 1147~Mhz.  The graphics card has 14 multiprocessors,
each with 32 cores, for a total of 448 cores.
As the clock speed of the GPU is a third of the clock speed of the CPU,
we hope to achieve a double digit speedup.
We used the NVIDIA CUDA compiler driver {\tt nvcc},
release 4.0, V0.2.1221.

In Table~\ref{tabexperiments1} and~\ref{tabexperiments2} 
we list results of our preliminary
implementation.  The number of threads in each block was~32 for
all three kernels to evaluate a system and its Jacobian matrix 
of dimension~32.  Generating 32 monomials per polynomial leads to
1,024 monomial in total.

\begin{table}[hbt]
\begin{center}
\begin{tabular}{r|r|r|r}
\#monomials & Tesla C2050 & 1 CPU core & speedup \\ \hline
 704 & 14.514~sec & 1min~50.9~sec &  7.60~~ \\
1024 & 15.265~sec & 2min~39.3~sec & 10.44~~ \\
1536 & 17.000~sec & 3min~58.7~sec & 14.04~~
\end{tabular}
\caption{Wall clock times and speedups for 100,000 evaluations 
of a polynomial system and its Jacobian matrix of dimension~32.
Each monomial has 9 variables occurring with nonzero power 
of at most 2.}
\label{tabexperiments1}
\end{center}
\end{table}

\begin{table}[hbt]
\begin{center}
\begin{tabular}{r|r|r|r}
\#monomials & Tesla C2050 & 1 CPU core & speedup \\ \hline
 704 & 19.068~sec & 3min~16.9~sec & 10.33~~ \\
1024 & 20.800~sec & 4min~43.3~sec & 13.62~~ \\
1536 & 21.763~sec & 7min~05.8~sec & 19.56~~
\end{tabular}
\caption{Wall clock times and speedups for 100,000 evaluations 
of a polynomial system and its Jacobian matrix of dimension~32.
Each monomial has 16 variables occurring with nonzero power 
of at most 10.}
\label{tabexperiments2}
\end{center}
\end{table}

Increasing the number of monomials to 2,048 in Table~\ref{tabexperiments1}
and~\ref{tabexperiments2}
would have yielded a speedup of more than~20, but the capacity of the
constant memory was not sufficient to hold the exponents and positions
of all 2,048 monomials.  For larger systems, we will upgrade our
preliminary implementation with a better compression strategy
(instead of the current {\tt char} used for each exponent).

\section{Conclusions}

Starting from an algorithm for evaluation and differentiation
that is already close to optimal, we obtained good speedups
on a graphics computing processor for randomly generated polynomial
systems of dimension 32 (the warp size) and fixed number of monomials
per polynomial in the system.

\bibliographystyle{plain}
% \bibliography{GPUpoleval}

\begin{thebibliography}{10}

\bibitem{Akl04}
S.G. Akl.
\newblock Superlinear performance in real-time parallel computation.
\newblock {\em The Journal of Supercomputing}, 29(1):89--111, 2004.

\bibitem{AG03}
E.L. Allgower and K.~Georg.
\newblock {\em Introduction to Numerical Continuation Methods}, volume~45 of
  {\em Classics in Applied Mathematics}.
\newblock SIAM, 2003.

\bibitem{ACW89}
D.C.S. Allison, A.~Chakraborty, and L.T. Watson.
\newblock Granularity issues for solving polynomial systems via globally
  convergent algorithms on a hypercube.
\newblock {\em J. of Supercomputing}, 3(1):5--20, 1989.

\bibitem{BHS11}
D.J. Bates, J.D. Hauenstein, and A.J. Sommese.
\newblock A parallel endgame.
\newblock In L.~Gurvits, P.~P{\'{e}}bay, J.M. Rojas, and D.~Thompson, editors,
  {\em Randomization, Relaxation, and Complexity in Polynomial Equation
  Solving}, volume 556 of {\em Contemporary Mathematics}, pages 25--35. AMS,
  2011.

\bibitem{BGKW08}
C.~Bischof, N.~Guertler, A.~Kowartz, and A.~Walther.
\newblock Parallel reverse mode automatic differentiation for {OpenMP} programs
  with {ADOL-C}.
\newblock In C.~Bischof, H.M. B{\"{u}}cker, P.~Hovland, U.~Naumann, and
  J.~Utke, editors, {\em Advances in Automatic Differentiation}, pages
  163--173. Springer-Verlag, 2008.

\bibitem{CARW93}
A.~Chakraborty, D.C.S. Allison, C.J. Ribbens, and L.T. Watson.
\newblock The parallel complexity of embedding algorithms for the solution of
  systems of nonlinear equations.
\newblock {\em IEEE Transactions on Parallel and Distributed Systems},
  4(4):458--465, 1993.

\bibitem{Dek71}
T.J. Dekker.
\newblock A floating-point technique for extending the available precision.
\newblock {\em Numerische Mathematik}, 18(3):224--242, 1971.

\bibitem{Eme09}
P.~Emeliyanenko.
\newblock Efficient multiplication of polynomials on graphics hardware.
\newblock In Y.~Dou, R.~Gruber, and J.M. Joller, editors, {\em Advanced
  Parallel Processing Technologies. 8th International Symposium, APPT 2009,
  Rapperswil, Switzerland, August 2009}, volume 5737 of {\em Lecture Notes in
  Computer Science}, pages 134--149. Springer-Verlag, 2009.

\bibitem{Eme10}
P.~Emeliyanenko.
\newblock A complete modular resultant algorithm targeted for realization on
  graphics hardware.
\newblock In M.M. Maza and J.-L. Roch, editors, {\em Proceedings of the 4th
  International Workshop on Parallel Symbolic Computation (PASCO 2010), July
  21-23 2010, Grenoble, France}, pages 35--43. ACM, 2010.

\bibitem{Eme11}
P.~Emeliyanenko.
\newblock High-performance polynomial {GCD} computations on graphics
  processors.
\newblock In W.W. Smari and J.P. McIntire, editors, {\em Proceedings of the
  2011 International Conference on High Performance Computing \mbox{\rm \&}
  Simulation (HPCS 2011)}, pages 215--224. IEEE, 2011.

\bibitem{GPGK08}
M.~Grabner, T.~Pock, T.~Gross, and B.~Kainz.
\newblock Automatic differentiation for {GPU}-accelerated {2D/3D} registration.
\newblock In C.~Bischof, H.M. B{\"{u}}cker, P.~Hovland, U.~Naumann, and
  J.~Utke, editors, {\em Advances in Automatic Differentiation}, pages
  259--269. Springer-Verlag, 2008.

\bibitem{GW08}
A.~Griewank and A.~Walther.
\newblock {\em Evaluating Derivatives: Principles and Techniques of Algorithmic
  Differentiation}.
\newblock SIAM, second edition, 2008.

\bibitem{GV08}
Y.~Guan and J.~Verschelde.
\newblock Parallel implementation of a subsystem-by-subsystem solver.
\newblock In {\em The proceedings of the 22th High Performance Computing
  Symposium, Quebec City, 9-11 June 2008}, pages 117--123. IEEE Computer
  Society, 2008.

\bibitem{GKFK06}
T.~Gunji, S.~Kim, K.~Fujisawa, and M.~Kojima.
\newblock {PHoMpara} -- parallel implementation of the \underline{P}olyhedral
  \underline{H}omotopy continuation \underline{M}ethod for polynomial systems.
\newblock {\em Computing}, 77(4):387--411, 2006.

\bibitem{HLB00}
Y.~Hida, X.S. Li, and D.H. Bailey.
\newblock Algorithms for quad-double precision floating point arithmetic.
\newblock In {\em {15th IEEE Symposium on Computer Arithmetic (Arith-15 2001),
  11-17 June 2001, Vail, CO, USA}}, pages 155--162. IEEE Computer Society,
  2001.
\newblock Shortened version of Technical Report LBNL-46996, software at {\tt
  http://crd.lbl.gov/$\sim$dhbailey/} {\tt mpdist/qd-2.3.9.tar.gz}.

\bibitem{Hwu11}
W.W. Hwu~(editor).
\newblock {\em GPU Computing Gems: Emerald Edition}.
\newblock Morgan Kaufmann, 2011.

\bibitem{KH10}
D.B. Kirk and W.W. Hwu.
\newblock {\em Programming Massively Parallel Processors. A Hands-on Approach}.
\newblock Morgan Kaufmann, 2010.

\bibitem{Koj08}
M.~Kojima.
\newblock Efficient evaluation of polynomials and their partial derivatives in
  homotopy continuation methods.
\newblock {\em Journal of the Operations Research Society of Japan},
  51(1):29--54, 2008.

\bibitem{Ley11}
A.~Leykin.
\newblock Numerical algebraic geometry.
\newblock {\em The Journal of Software for Algebra and Geometry: Macaulay2},
  3:5--10, 2011.

\bibitem{LV05}
A.~Leykin and J.~Verschelde.
\newblock Factoring solution sets of polynomial systems in parallel.
\newblock In T.~Skeie and C.-S. Yang, editors, {\em Proceedings of the 2005
  International Conference on Parallel Processing Workshops. 14-17 June 2005.
  Oslo, Norway. High Performance Scientific and Engineering Computing}, pages
  173--180. IEEE Computer Society, 2005.

\bibitem{LV09}
A.~Leykin and J.~Verschelde.
\newblock Decomposing solution sets of polynomial systems: a new parallel
  monodromy breakup algorithm.
\newblock {\em The International Journal of Computational Science and
  Engineering}, 4(2):94--101, 2009.

\bibitem{LVZh06}
A.~Leykin, J.~Verschelde, and Y.~Zhuang.
\newblock Parallel homotopy algorithms to solve polynomial systems.
\newblock In N.~Takayama and A.~Iglesias, editors, {\em Proceedings of ICMS
  2006}, volume 4151 of {\em Lecture Notes in Computer Science}, pages
  225--234. Springer-Verlag, 2006.

\bibitem{Li03}
T.Y. Li.
\newblock Numerical solution of polynomial systems by homotopy continuation
  methods.
\newblock In F.~Cucker, editor, {\em Handbook of Numerical Analysis. Volume XI.
  Special Volume: Foundations of Computational Mathematics}, pages 209--304.
  North-Holland, 2003.

\bibitem{LT09}
T.Y. Li and C.-H. Tsai.
\newblock {HOM4PS-2.0para}: Parallelization of {HOM4PS-2.0} for solving
  polynomial systems.
\newblock {\em Parallel Computing}, 35(4):226--238, 2009.

\bibitem{LHL10}
M.~Lu, B.~He, and Q.~Luo.
\newblock Supporting extended precision on graphics processors.
\newblock In A.~Ailamaki and P.A. Boncz, editors, {\em Proceedings of the Sixth
  International Workshop on Data Management on New Hardware (DaMoN 2010), June
  7, 2010, Indianapolis, Indiana}, pages 19--26, 2010.
\newblock Software at {\tt http://code.google.com/p/gpugrpec/}.

\bibitem{MP11}
M.M. Maza and W.~Pan.
\newblock Solving bivariate polynomial systems on a {GPU}.
\newblock In HPCS 2011, Montreal, 15-17 June 2011, to appear in Journal of
  Physics: Conference Series.

\bibitem{MP10}
M.M. Maza and W.~Pan.
\newblock Fast polynomial multiplication on a {GPU}.
\newblock {\em Journal of Physics: Conference Series}, 256, 2010.
\newblock High Performance Computing Symposium (HPCS2010), 5-9 June 2010,
  Victoria College, University of Toronto, Canada.

\bibitem{Mor87}
A.~Morgan.
\newblock {\em Solving polynomial systems using continuation for engineering
  and scientific problems}.
\newblock Prentice-Hall, 1987.
\newblock Volume 57 of Classics in Applied Mathematics Series, SIAM 2009.

\bibitem{NVIDIA}
NVIDIA.
\newblock {NVIDIA} {CUDA} {P}rogramming {G}uide. {V}ersion 3.0.
\newblock 2010.

\bibitem{Pri92}
D.N. Priest.
\newblock {\em On Properties of Floating Point Arithmetics: Numerical Stability
  and the Cost of Accurate Computations}.
\newblock PhD thesis, University of California at Berkeley, 1992.
\newblock {\tt ftp://ftp.icsi.berkeley.edu/pub/theory/} {\tt
  priest-thesis.ps.Z}.

\bibitem{Rum10}
S.M. Rump.
\newblock Verification methods: Rigorous results using floating-point
  arithmetic.
\newblock {\em Acta Numerica}, 19:287–449, 2010.

\bibitem{She97}
J.R. Shewchuk.
\newblock Adaptive precision floating-point arithmetic and fast robust
  geometric predicates.
\newblock {\em Discrete Comput.\ Geom.}, 18(3):305--363, 1997.

\bibitem{MPI98}
M.~Snir, S.~Otto, S.~Huss-Lederman, D.~Walker, and J.~Dongarra.
\newblock {\em {MPI - The Complete Reference Volume 1, The MPI Core}}.
\newblock Massachusetts Institute of Technology, second edition, 1998.

\bibitem{SVW05}
A.J. Sommese, J.~Verschelde, and C.W. Wampler.
\newblock Introduction to numerical algebraic geometry.
\newblock In A.~Dickenstein and I.Z. Emiris, editors, {\em Solving Polynomial
  Equations. Foundations, Algorithms and Applications}, volume~14 of {\em
  Algorithms and Computation in Mathematics}, pages 301--337. Springer-Verlag,
  2005.

\bibitem{SW05}
A.J. Sommese and C.W. Wampler.
\newblock {\em The Numerical solution of systems of polynomials arising in
  engineering and science}.
\newblock World Scientific, 2005.

\bibitem{SMSW06}
H.-J. Su, J.M. McCarthy, M.~Sosonkina, and L.T. Watson.
\newblock Algorithm 857: {POLSYS\_GLP}: A parallel general linear product
  homotopy code for solving polynomial systems of equations.
\newblock {\em ACM Trans. Math. Softw.}, 32(4):561--579, 2006.

\bibitem{Ver99}
J.~Verschelde.
\newblock Algorithm 795: {PHC}pack: A general-purpose solver for polynomial
  systems by homotopy continuation.
\newblock {\em ACM Trans. Math. Softw.}, 25(2):251--276, 1999.
\newblock Software available at {\tt
  http://www.math.uic.edu/{\~{}}jan/download.html}.

\bibitem{VW04}
J.~Verschelde and Y.~Wang.
\newblock Computing feedback laws for linear systems with a parallel {P}ieri
  homotopy.
\newblock In Y.~Yang, editor, {\em Proceedings of the 2004 International
  Conference on Parallel Processing Workshops, 15-18 August 2004, Montreal,
  Quebec, Canada. High Performance Scientific and Engineering Computing}, pages
  222--229. IEEE Computer Society, 2004.

\bibitem{VY11}
J.~Verschelde and G.~Yoffe.
\newblock Quality up in polynomial homotopy continuation by multithreaded path
  tracking.
\newblock Preprint {\tt arXiv:1109.0545v1 [cs.DC] 2 Sep 2011}.

\bibitem{VY10}
J.~Verschelde and G.~Yoffe.
\newblock Polynomial homotopies on multicore workstations.
\newblock In M.M. Maza and J.-L. Roch, editors, {\em Proceedings of the 4th
  International Workshop on Parallel Symbolic Computation (PASCO 2010), July
  21-23 2010, Grenoble, France}, pages 131--140. ACM, 2010.

\bibitem{VZ06}
J.~Verschelde and Y.~Zhuang.
\newblock Parallel implementation of the polyhedral homotopy method.
\newblock In T.M. Pinkston and F.~Ozguner, editors, {\em Proceedings of the
  2006 International Conference on Parallel Processing Workshops. 14-18
  Augustus 2006. Columbus, Ohio. High Performance Scientific and Engineering
  Computing}, pages 481--488. IEEE Computer Society, 2006.

\bibitem{Wat02}
L.T. Watson.
\newblock Probability-one homotopies in computational science.
\newblock {\em J. Comput. Appl. Math.}, 140(1\&2):785--807, 2002.

\end{thebibliography}

\end{document}